\def\year{2017}
\documentclass[letterpaper]{article}
\usepackage{aaai17}
\usepackage{subfig}
\usepackage{hyperref}

\usepackage{xspace}
\usepackage{multirow} \usepackage{siunitx} \usepackage{booktabs} 
\newcommand{\usergraph}{{\ensuremath {\cal G}}\xspace}
\newcommand{\tweetsreplies}{\ensuremath{ {\cal T}}\xspace}
\newcommand{\usersreplies}{\ensuremath{ {\cal R}}\xspace}

\newcommand{\spara}[1]{\smallskip\noindent\textbf{#1}}

\usepackage{tikz,pgfplots,pgfplotstable}
\usetikzlibrary{positioning,fit,shapes}
\usetikzlibrary{decorations.pathreplacing}
\usetikzlibrary{arrows}

\pgfdeclarelayer{background}
\pgfdeclarelayer{foreground}
\pgfsetlayers{background,main,foreground}

\makeatletter
\tikzset{multicircle/.style  args={#1, #2}{ alias=tmp@name,   postaction={    insert path={
     \pgfextra{     \pgfpointdiff{\pgfpointanchor{\pgf@node@name}{center}}                  {\pgfpointanchor{\pgf@node@name}{east}}     \pgfmathsetmacro\insiderad{\pgf@x}             \fill[white] (\pgf@node@name.center)  circle (\insiderad-\pgflinewidth);        \draw[#2] (\pgf@node@name.center)  circle (\insiderad-\pgflinewidth);        \fill[#2] (\pgf@node@name.center)  -- ++(0:\insiderad-\pgflinewidth) arc (0:#1:\insiderad-\pgflinewidth)--cycle;        }}}}}
\makeatother

\definecolor{yafaxiscolor}{rgb}{0.3, 0.3, 0.3}
\definecolor{yafcolor1}{rgb}{0.4, 0.165, 0.553}
\definecolor{yafcolor2}{rgb}{0.949, 0.482, 0.216}
\definecolor{yafcolor3}{rgb}{0.47, 0.549, 0.306}
\definecolor{yafcolor4}{rgb}{0.925, 0.165, 0.224}
\definecolor{yafcolor5}{rgb}{0.141, 0.345, 0.643}
\definecolor{yafcolor6}{rgb}{0.965, 0.933, 0.267}
\definecolor{yafcolor7}{rgb}{0.627, 0.118, 0.165}
\definecolor{yafcolor8}{rgb}{0.878, 0.475, 0.686}
\definecolor{yafcolor9}{rgb}{0.965, 0.733, 0.767}

\newlength{\yafaxispad}
\newlength{\yaftlpad}
\newlength{\yaflabelpad}
\newlength{\yafaxiswidth}
\newlength{\yafticklen}
\makeatletter
\def\pgfplots@drawtickgridlines@INSTALLCLIP@onorientedsurf#1{}
\makeatother

\newcommand{\yafdrawxaxis}[2]{
  \pgfplotstransformcoordinatex{#1}\let\xmincoord=\pgfmathresult 
  \pgfplotstransformcoordinatex{#2}\let\xmaxcoord=\pgfmathresult 
  \pgfsetlinewidth{\yafaxiswidth} 
  \pgfsetcolor{yafaxiscolor}
  \pgfpathmoveto{\pgfpointadd{\pgfpointadd{\pgfplotspointrelaxisxy{0}{0}}{\pgfqpointxy{\xmincoord}{0}}}{\pgfqpoint{-0.5\yafaxiswidth}{\yafaxispad}}}
  \pgfpathlineto{\pgfpointadd{\pgfpointadd{\pgfplotspointrelaxisxy{0}{0}}{\pgfqpointxy{\xmaxcoord}{0}}}{\pgfqpoint{0.5\yafaxiswidth}{\yafaxispad}}}
  \pgfusepath{stroke}

}
\newcommand{\yafdrawyaxis}[2]{
  \pgfplotstransformcoordinatey{#1}\let\ymincoord=\pgfmathresult 
  \pgfplotstransformcoordinatey{#2}\let\ymaxcoord=\pgfmathresult 
  \pgfsetlinewidth{\yafaxiswidth} 
  \pgfsetcolor{yafaxiscolor}
  \pgfpathmoveto{\pgfpointadd{\pgfpointadd{\pgfplotspointrelaxisxy{0}{0}}{\pgfqpointxy{0}{\ymincoord}}}{\pgfqpoint{\yafaxispad}{-0.5\yafaxiswidth}}}
  \pgfpathlineto{\pgfpointadd{\pgfpointadd{\pgfplotspointrelaxisxy{0}{0}}{\pgfqpointxy{0}{\ymaxcoord}}}{\pgfqpoint{\yafaxispad}{0.5\yafaxiswidth}}}
  \pgfusepath{stroke}
}

\pgfplotscreateplotcyclelist{yaf}{{yafcolor1,mark options={scale=0.75},mark=o}, 
{yafcolor2,mark options={scale=0.75},mark=square},
{yafcolor3,mark options={scale=0.75},mark=triangle},
{yafcolor4,mark options={scale=0.75},mark=o},
{yafcolor5,mark options={scale=0.75},mark=o},
{yafcolor6,mark options={scale=0.75},mark=o},
{yafcolor7,mark options={scale=0.75},mark=o},
{yafcolor8,mark options={scale=0.75},mark=o}} 

\pgfplotsset{axis y line=left, axis x line=bottom,
  tick align=outside,
  compat = 1.3,
  tickwidth=\yafticklen,
  clip = false,
  every axis title shift = 0pt,
    x axis line style= {-, line width = 0pt, opacity = 0},
    y axis line style= {-, line width = 0pt, opacity = 0},
    x tick style= {line width = \yafaxiswidth, color=yafaxiscolor, yshift = \yafaxispad},
    y tick style= {line width = \yafaxiswidth, color=yafaxiscolor, xshift = \yafaxispad},
    x tick label style = {font=\scriptsize, yshift = \yaftlpad},
    y tick label style = {font=\scriptsize, xshift = \yaftlpad},
    every axis y label/.style = {at = {(ticklabel cs:0.5)}, rotate=90, anchor=center, font=\scriptsize, yshift = -\yaflabelpad},
    every axis x label/.style = {at = {(ticklabel cs:0.5)}, anchor=center, font=\scriptsize, yshift = \yaflabelpad},
    x tick label style = {font=\scriptsize, yshift = 1pt},
    grid = major,
    major grid style  = {dash pattern = on 1pt off 3 pt},
  every axis plot post/.append style= {line width=\yafaxiswidth} ,
  legend cell align = left,
  legend style = {inner sep = 1pt, cells = {font=\scriptsize}},
  legend image code/.code={    \draw[mark repeat=2,mark phase=2,#1] 
    plot coordinates { (0cm,0cm) (0.15cm,0cm) (0.3cm,0cm) };  } 
}

\begin{document}

\title{A Motif-based Approach for Identifying Controversy}

\author{Mauro Coletto\\Ca' Foscari Uni. - Venice\\\textit{mauro.coletto@unive.it} \And
Kiran Garimella\\Aalto University - Helsinki\\\textit{kiran.garimella@aalto.fi} \And
Aristides Gionis\\Aalto University - Helsinki\\\textit{aristides.gionis@aalto.fi}\And
Claudio Lucchese\\CNR Pisa\\\textit{claudio.lucchese@isti.cnr.it}  }

\maketitle

\begin{abstract}
Among the topics discussed in Social Media, 
some lead to controversy. 
A number of recent studies have focused 
on the problem of identifying controversy in social media
mostly 
based on the analysis of textual content
or rely on global network structure.
Such approaches have strong limitations due to the difficulty of understanding natural language, 
and of investigating the global network structure.

In this work we show that it is possible to detect controversy in social media by exploiting network motifs, 
i.e., local patterns of user interaction. 
The proposed approach allows for a language-independent
and fine-grained and efficient-to-compute analysis of user discussions and their evolution over time.
The supervised model exploiting motif patterns can achieve 85\% accuracy, 
with an improvement of 7\% compared to baseline structural, propagation-based and temporal network features.

\end{abstract}

\section{Introduction}
\label{sec:intro}

In this paper we study the problem of identifying controversies in social media, which has recently drawn some attention~\cite{garimella2016quantifying,coletto2016polarized}. 
However, as this is a difficult problem, 
involving processing of human language and network dynamics, 
existing studies have limitations.
For example, many papers 
study controversy in very controlled case studies, 
or focus on a predefined topic, most typically politics~\cite{conover2011political}, 
for which they employ auxiliary domain-specific sources and datasets.
In other cases, proposed approaches are based on content-based analysis~\cite{mejova2014controversy}, 
which has several limitations, as well, 
due to the ambiguity of the language and the fact that models 
become language-dependent and topic-dependent.
We aim to identify controversies
on {\em any} topic, 
discussed in {\em any} language.
In this sense, 
our paper is related to the recent work of Garimella et al.~\cite{garimella2016quantifying}, 
who also aim at identifying controversies based on the 
analysis of the {\em network structure}. 
An obvious limitation in their work is that they assume that a topic partitions the network always into two clusters 
and that it is computationally feasible to identify those clusters. 
In our work, we overcome those limitations
by analyzing local network patterns ({\em motifs}), 
and thus, making no assumption about the global cluster structure of the network, 
or about our ability to detect network clusters. 
Moreover, note that the separation of the retweet network in communities 
does not always reflect controversy;
it may also mean that a hashtag is used in two communities with different acceptations. Our approach catches antagonism in the conversation and it allows to dynamically discover potential controversial sub-discussions
that may be present within an otherwise non-controversial topic.

\section{Data collection}
\label{sec:data}

\spara{Dataset: {\em Twitter pages.}} Our main source of data is a carefully-curated set of popular Twitter pages which covers a wide range of domains 
(news, politics, celebrity, gossip, entertainment) and languages. 
For each page, we gather the last two hundreds tweets and we manually evaluate them to check if they are controversial or not through multiple annotators. To classify them the content of the tweet and the received replies were considered. A tweet is labeled controversial if the content is debatable and it expresses an idea or an opinion which generates an argument in the replies, representing opposing opinions in favor or in disagreement with the root tweet.
We consider only the pages whose tweets are almost completely controversial or not controversial resulting in 11 controversial and 7 non-controversial pages: a tweet is deemed controversial (non-controversial) if it originates from a controversial (non-controversial) classified page. For each collected tweet in each page (\textit{root post}), we reconstructed the generated discussion thread
by recursively crawling the tweet's replies. We restrict to the tweets that generate a conversation involving more than $k$ users, with k=2,3 and 10. (including the author of the original post). 
Table~\ref{tab:data2} reports the number of root posts and total reply tweets that we collect. The final dataset contains more than $190K$ tweets in total.  
Each collected root post generates a network of replies that involves on average about 100 users.

\begin{table}[t]
\caption{Dataset Statistics.}
\centering
\scriptsize
\label{tab:data2} 
\begin{tabular}{llcc}
\multicolumn{4}{c}{\textbf{{\em Twitter pages}}} \\
\toprule
Filtering & \multicolumn{1}{c}{{Root posts}} & {Avg. users} & {Tot. tweets}\\ 
\hline
$>$2 users & \num{1202} & \num{108} & $192.7 K$\\
$>$3 users & \num{1175} (97\%)  & \num{110} & $192.5 K$\\
$>$10 users & \num{1046} (87\%) & \num{123} & $191.3 K$\\
\hline
\end{tabular}
\end{table} 
\section{Controversy analysis and detection}
\label{sec:method}

Given a social network
we are interested in modeling the interactions among users
and the dynamics incurring due to generated content.
We consider a {\em user graph} $\usergraph=(U,E)$,
where $U$ is the set of users of the network
and an edge $e = (u_i,u_j) \in E$ indicates that 
user $u_i$ follows user $u_j$.
Moreover, a user may publish some new content item $c_i$, 
possibly {\em in response to} another content item $c_j$
authored by another user, 
thus generating complex threads of discussion. 
Interactions within a single thread are modeled with a content {\em reply tree} 
$\tweetsreplies=(C,R)$, 
where $C$ is the set of content items in the thread, 
and an arc $r=(c_i,c_j)\in R$ indicates that $c_i$ is a reply to $c_j$. 
The tree $\tweetsreplies$ can be projected onto the users
to model reply interactions among users.
The resulting structure is a user {\em reply graph} 
$\usersreplies=(U,I)$, 
where an edge $e=(u_i,u_j)\in I$ indicates that the user $u_i$ has replied to some 
content item posted by user $u_j$.
Our hypothesis
is that the structure of \usergraph, \tweetsreplies, and \usersreplies 
can be characterized by simple {\em motifs} of local user interactions useful to distinguish between {\em controversial} and {\em non-controversial} content.
In addition to local motifs, 
we also explore whether baseline features
(including network structure, content propagation, and temporal features)
are predictors of controversy.

\subsection{Graph-based analysis}
\label{section:standard-features}

\spara{Structural features.}
The simplest structural features to extract from the user-interaction networks
are the {\em size} in terms of {\em number of nodes} and {\em number of edges}, 
and the {\em degree distribution}.

\begin{figure*}[t]
\centering
\subfloat[][Frequency]{\includegraphics[height=3.5cm]{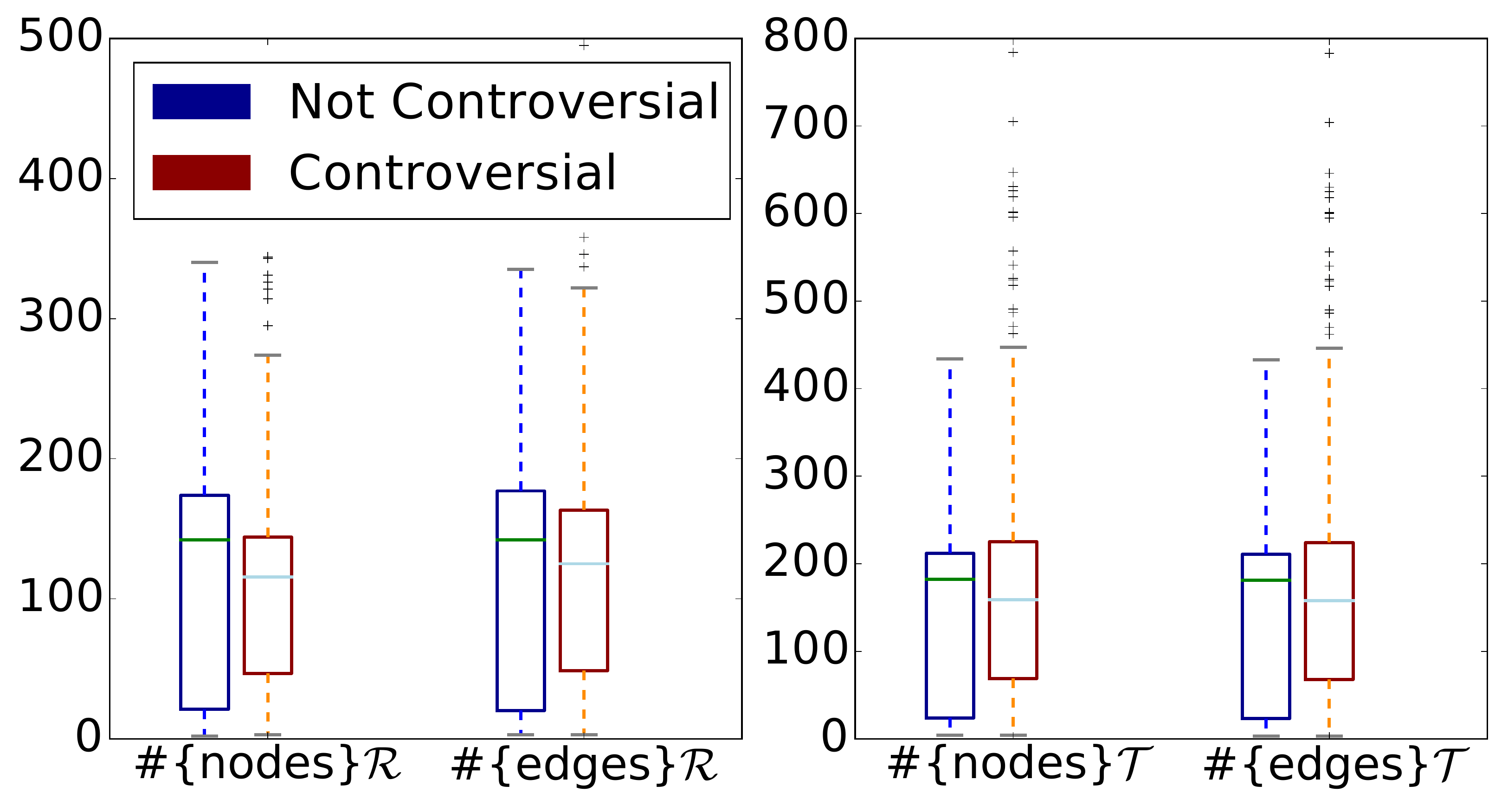}\label{fig:pop}}
\subfloat[][Structural]{\includegraphics[height=3.4cm]{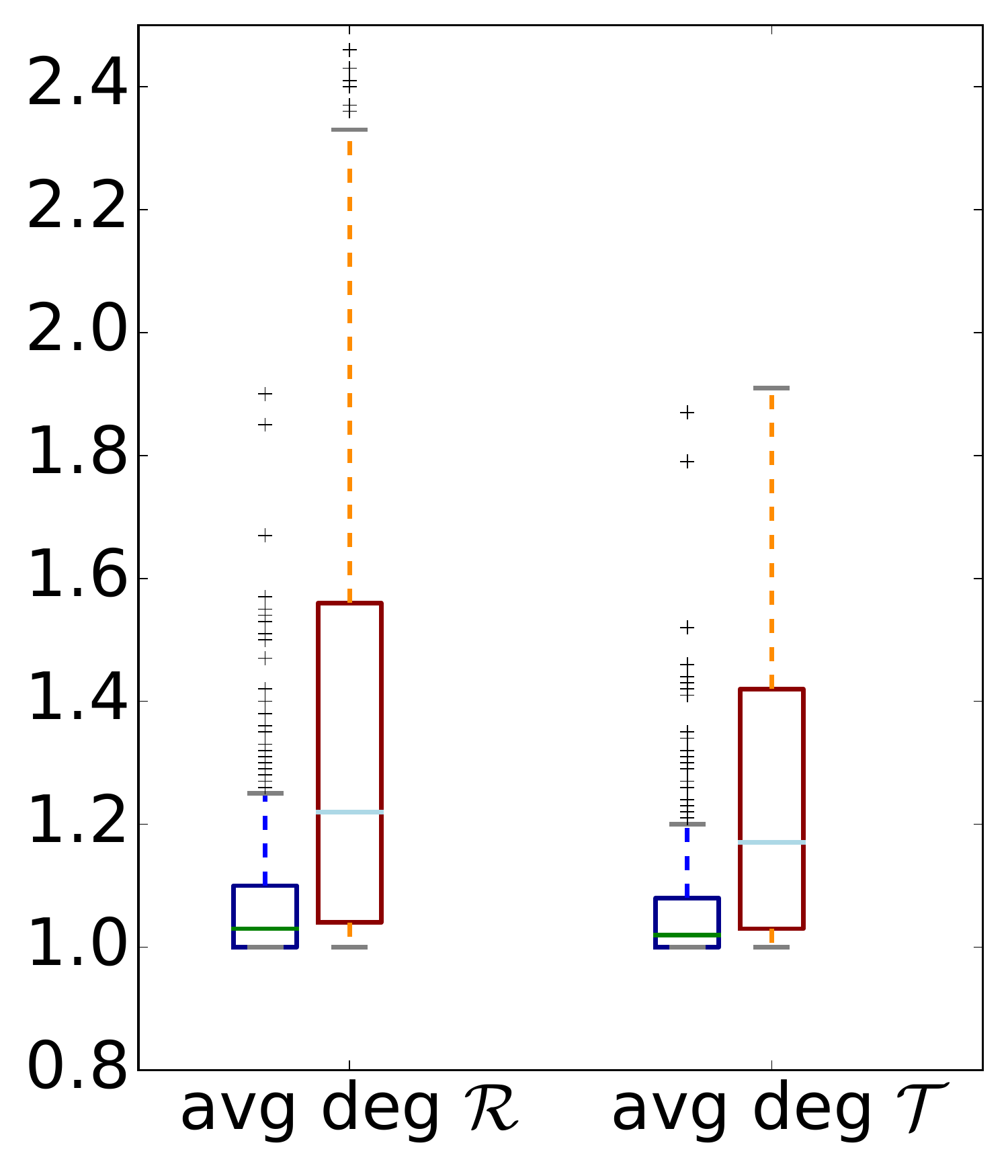}\label{fig:struct}}
\subfloat[][Propagation I]{\includegraphics[height=3.5cm]{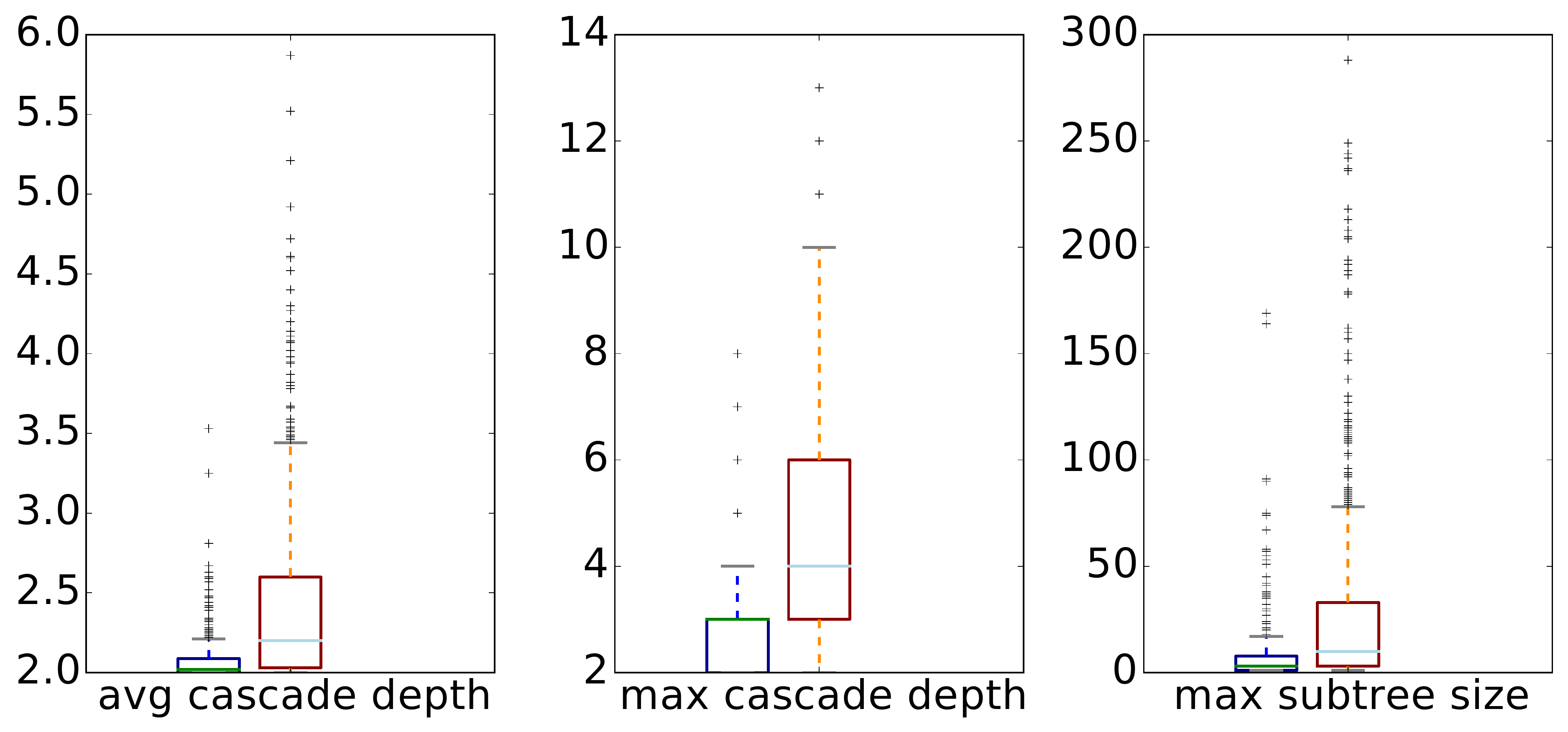}\label{fig:prop1}}\\
\subfloat[][Propagation II]{\includegraphics[height=3.5cm]{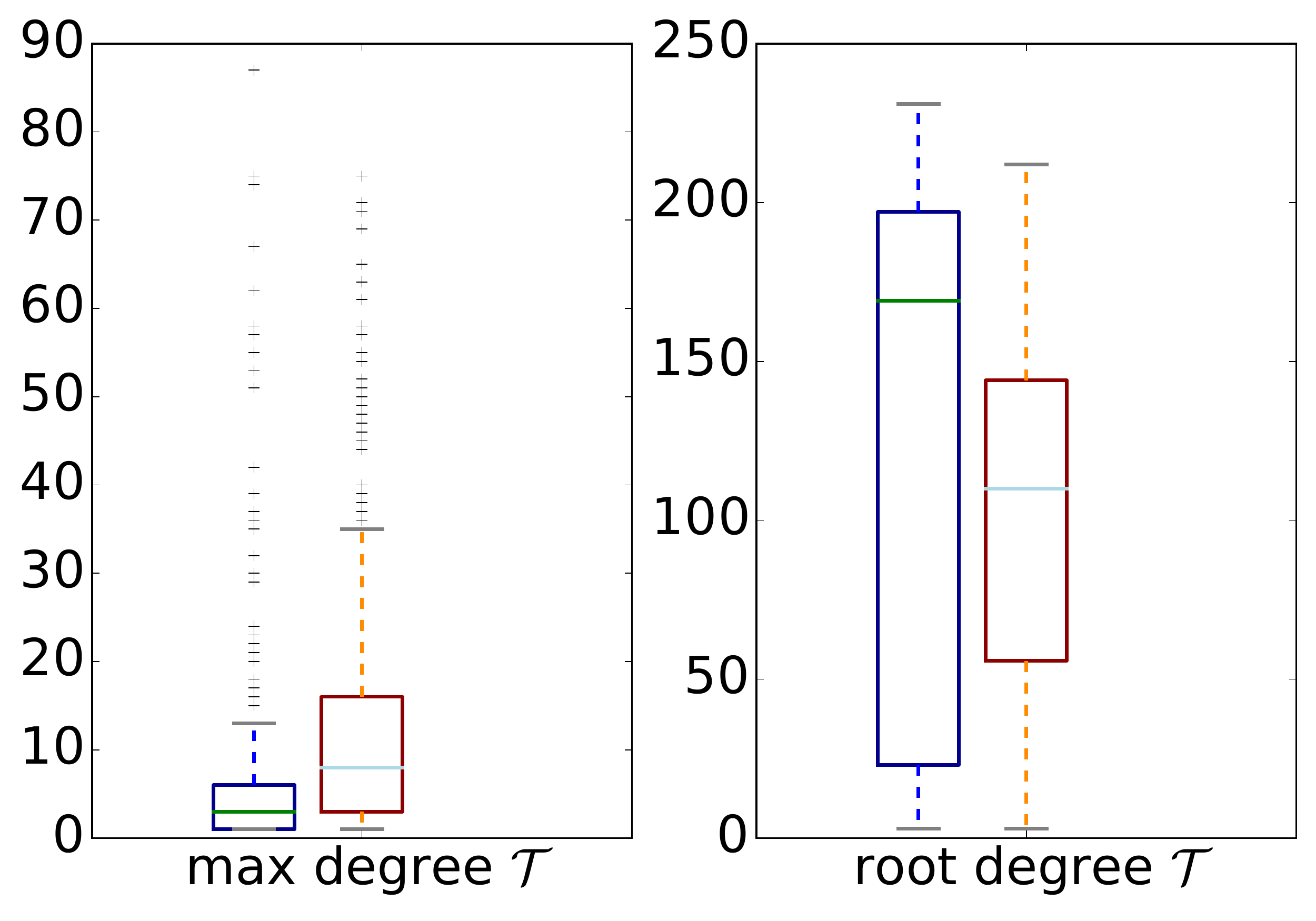}\label{fig:prop2}}
\subfloat[][Time]{\includegraphics[height=3.5cm]{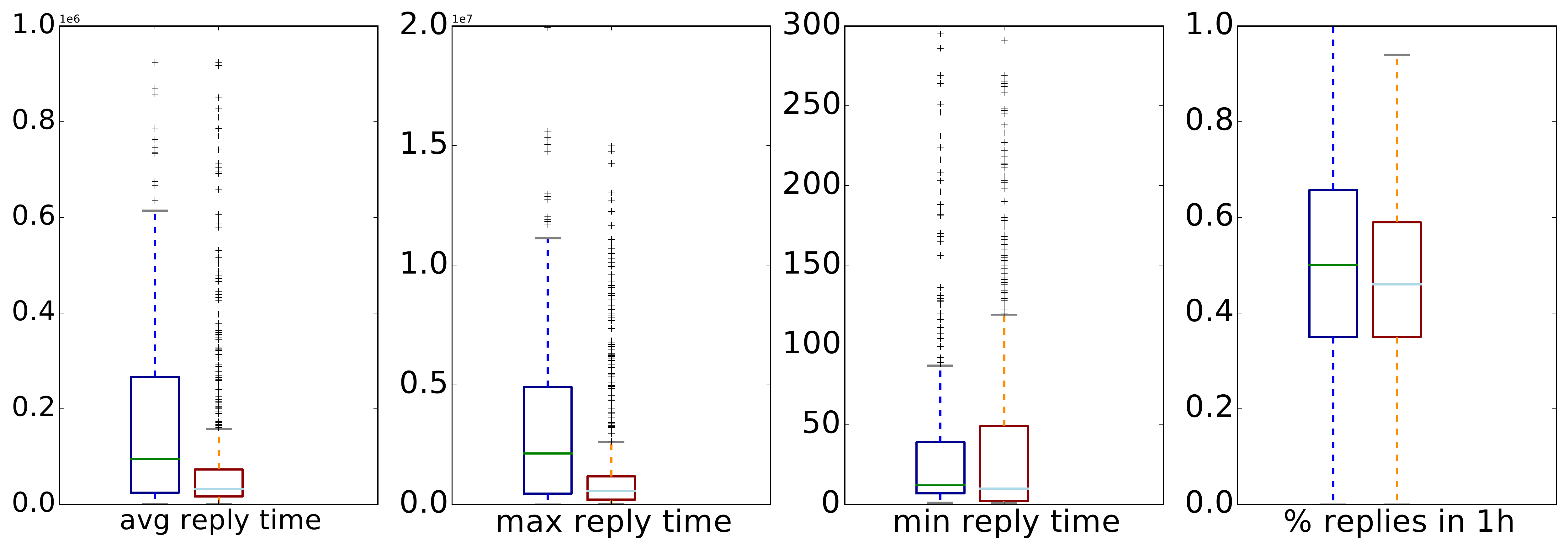}\label{fig:time}}
\caption{(a) Distribution of the number of nodes and edges in \tweetsreplies and \usersreplies. 
(b) Distribution of average node degree in \tweetsreplies and \usersreplies.
(c) Distribution of avg./max. cascade depth and max. subtree size.
(d) Distribution of origin degree and max. degree in \tweetsreplies and \usersreplies.
(e) Distribution of average, max., min. inter-reply time, and percentage of replies within one hour from the root.
 Non-controversial in blue (left side)  vs.\ controversial in red (right side).
 }
\label{fig:structural}
\end{figure*}

Figure~\ref{fig:pop} shows the distribution of the sizes of the 
reply tree \tweetsreplies and the reply graph \usersreplies
in terms of number of nodes and number of edges in the dataset (at least 3 users involved in the conversation).
Note that in our data the sizes of \tweetsreplies and \usersreplies are very similar 
for both controversial and non-controversial content.
This finding is in line with Smith et al.~\cite{smith2013role}
that controversial content does not necessarily generate larger threads of conversation.

Figure~\ref{fig:struct} reports the average degree for 
the reply tree \tweetsreplies and the reply graph \usersreplies.
In this case, the distributions are quite different: a larger average degree is observed for controversial content, suggesting that such conversations
generate more engagement among users. 

\spara{Propagation-based features.}
In order to understand how information propagates,
we investigate a number of different properties of the reply trees \tweetsreplies related to information propagation. 

Figure~\ref{fig:prop1} shows the distribution of average and maximum cascade depths,
where a cascade is defined as a path from the root to a leaf of a reply tree.
The figure also shows the distribution of the maximum-size subtree
among all subtrees rooted in a child of the root node.
We observe that for controversial content the reply trees generally have larger depth. 

Figure~\ref{fig:prop2} reports the distribution of the degree
for the root, as well as the node with the larger degree excluding the root in \tweetsreplies.
Reply trees of controversial discussions have higher probability of having a smaller root degree 
than non-controversial, suggesting that controversial discussions go beyond the first level of interaction.
We decided to use the two most significant features in the content reply trees: (\em average cascade depth) the average length of root-to-leaf paths and (\em maximum relative degree) the largest node degree excluding the root node, divided by the degree of the root. The other features, e.g.\ max cascade depth, are discarded because they are strongly related to popularity. 

\spara{Temporal features.}
Considering the simple assumption that controversial topics may generate ``dense'' discussions in time, 
we analyze the time elapsed between a content item and its reply (Figure~\ref{fig:time}).
Additionally, we measure the ratio of nodes in a reply tree occurring within one hour from the root. 
For prediction purposes, we chose to use as features only the average inter-reply time and the ratio of replies
in the first hour, since maximum and minimum inter-reply time are influenced by a single reply.

\begin{figure*}
\begin{center}
\subfloat[][]{\begin{tikzpicture} [scale=0.7,every node/.style={scale=0.7}]

\tikzstyle{vertex} = [inner sep = 1pt,  minimum width=11pt, 
	thick, circle, text=black!90, draw=yafcolor1!90, fill=yafcolor1!05]
\tikzstyle{line} = [thick]
\tikzstyle{plaintext} = [font=\large, text centered, text=black!90]

\tikzset{
  text style/.style={text=black!90}
}

\node[vertex] (au) at (1,4) {};
\node[vertex] (av) at (3,4) {};
\node[plaintext] (A) at (0.3,4) {A};

\node[vertex] (bu) at (1,3) {};
\node[vertex] (bv) at (3,3) {};
\node[plaintext] (B) at (0.3,3) {B};

\node[vertex] (cu) at (5,4) {};
\node[vertex] (cv) at (7,4) {};
\node[plaintext] (C) at (4.3,4) {C};

\node[vertex] (du) at (5,3) {};
\node[vertex] (dv) at (7,3) {};
\node[plaintext] (D) at (4.3,3) {D};

\node[vertex] (eu) at (9,4) {};
\node[vertex] (ev) at (11,4) {};
\node[plaintext] (E) at (8.3,4) {E};

\node[vertex] (fu) at (9,3) {};
\node[vertex] (fv) at (11,3) {};
\node[plaintext] (F) at (8.3,3) {F};

\node[vertex] (gu) at (9,2) {};
\node[vertex] (gv) at (11,2) {};
\node[plaintext] (G) at (8.3,2) {G};

\draw (au) edge [yafcolor3!90, line, ->] (av);

\draw (bu) edge [yafcolor3!90, line, <->] (bv);

\path (cu) edge [yafcolor3!90, line, ->, bend left = 10, transform canvas={yshift=1pt}] (cv);
\path (cu) edge [yafcolor2!90, line, ->, bend right = 10, transform canvas={yshift=-1pt}] (cv);

\path (du) edge [yafcolor3!90, line, ->, bend left = 10, transform canvas={yshift=1pt}] (dv);
\path (du) edge [yafcolor2!90, line, <-, bend right = 10, transform canvas={yshift=-1pt}] (dv);

\path (eu) edge [yafcolor3!90, line, <->, bend left = 10, transform canvas={yshift=1pt}] (ev);
\path (eu) edge [yafcolor2!90, line, <->, bend right = 10, transform canvas={yshift=-1pt}] (ev);

\path (fu) edge [yafcolor3!90, line, <->, bend left = 10, transform canvas={yshift=1pt}] (fv);
\path (fu) edge [yafcolor2!90, line, <-, bend right = 10, transform canvas={yshift=-1pt}] (fv);

\path (gu) edge [yafcolor3!90, line, ->, bend left = 10, transform canvas={yshift=1pt}] (gv);
\path (gu) edge [yafcolor2!90, line, <->, bend right = 10, transform canvas={yshift=-1pt}] (gv);

\node[vertex] (xr) at (2,1.5) {};
\node[vertex] (yr) at (4,1.5) {};
\draw (xr) edge [yafcolor3!90, line, ->] node [text style, above, yshift=1pt] {replies to} (yr);

\node[vertex] (xf) at (2,0.7) {};
\node[vertex] (yf) at (4,0.7) {};
\draw (xf) edge [yafcolor2!90, line, ->] node [text style, above,yshift=1pt] {follows} (yf);

\end{tikzpicture} \label{figure:motifs-2}}\hspace{1cm}
\subfloat[][]{\includegraphics[clip=true,width=\columnwidth]{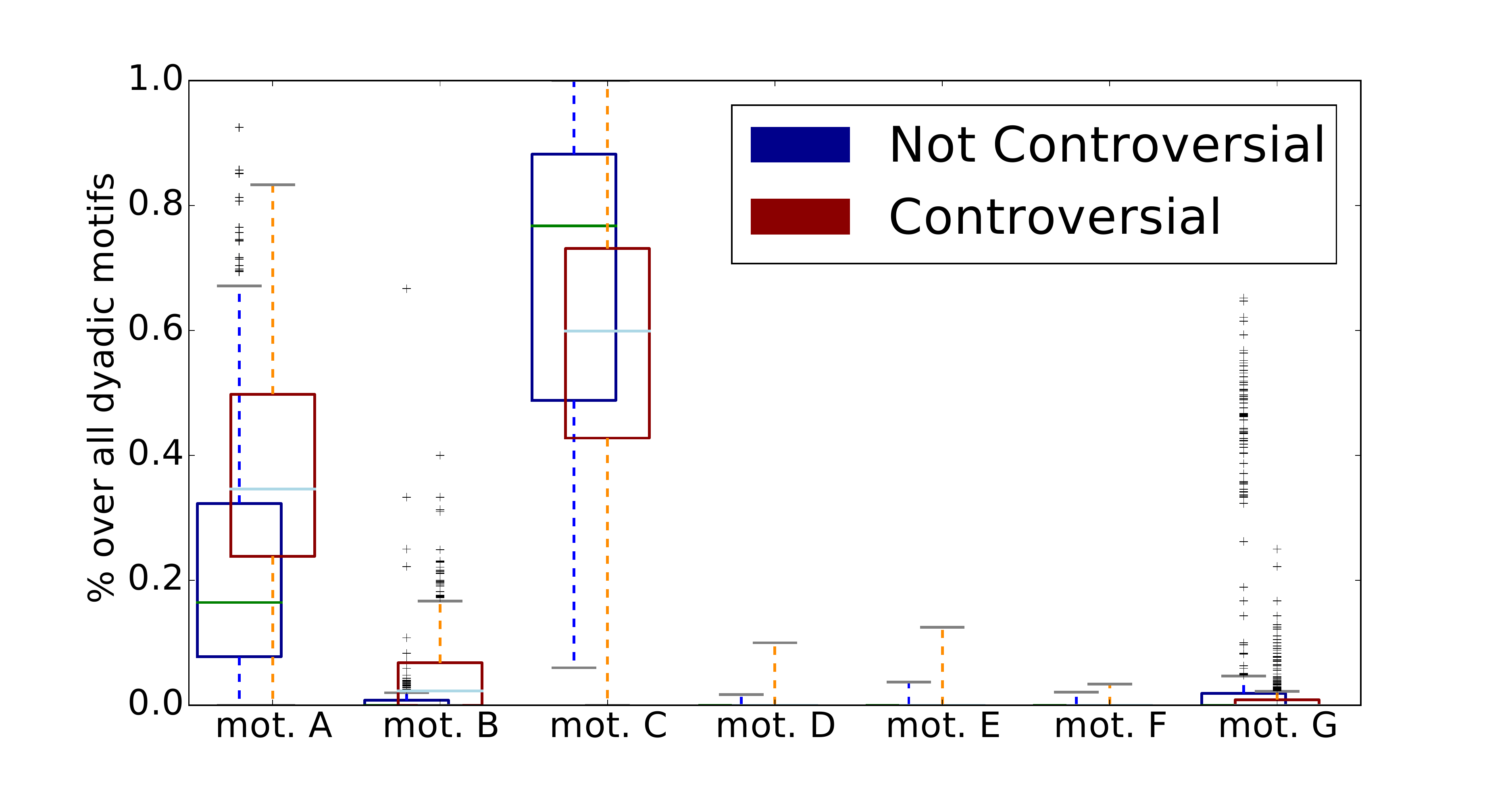} \label{fig:motifdata}}
\caption{(a) Dyadic motifs and (b) their frequency distribution.}
\end{center}
\end{figure*}

\spara{Motifs}
Our main hypothesis in this paper is that {\em local patterns} of 
user interaction can be used to discriminate between
controversial and non-controversial discussions.
This hypothesis is consistent with previous studies, 
where it was shown that local patterns can be used to characterize different types of networks~\cite{milo2002network}.
We consider motifs in the user graph \usergraph and the reply graph \usersreplies.
An edge in the user graph \usergraph indicates that a user follows another user. 
These two users are likely to have similar interests and/or opinions. 
On the other hand, the reply graph \usersreplies models the activity among users who may not know each other 
but they are willing to discuss or comment on a specific topic.
In this sense, the reply graph \usersreplies is much more dynamic and content-dependent.
Antagonism between users, which can not be captured by the user graph \usergraph can be captured by the reply graph \usersreplies.
Our basic assumption is that a combined analysis of the two graphs, 
\usergraph and  \usersreplies, 
can lead to an improved model for controversy detection.

We consider all possible patterns between two users in graphs \usergraph and \usersreplies, 
such that that there is at least one reply.
There are seven possible configurations (Figure~\ref{figure:motifs-2}).
Figure~\ref{fig:motifdata} shows the frequency distribution of dyadic motifs in our data. Note that patterns are mutually exclusive.
The most frequent dyadic motifs are $A$ and $C$. 
According to Figure~\ref{fig:motifdata},
it is more likely to observe a reply to a followed user in non-controversial cases.
Conversely, in controversial cases it is likely to reply to a user not being followed, confirming our intuitions.
The features used
for detecting controversial content
are the frequencies of all dyadic motifs.

We also consider 3-node motifs, in particular closed triangles. 
As in the case of dyadic motifs, 
we combine structural information from the user graph \usersreplies and the reply graph \usergraph.
Due to the high number of possible motifs
and since most motifs are relatively rare in the data, 
we coalesce motifs in groups (20).
The frequency of each group is considered as a feature for predicting controversy.
For the lack of space we do not report the distribution for all the motifs, but generally most of the patterns we considered for closed triangles were quite rare in the dataset.
Only a few of them are frequent and mostly in controversial threads, 
confirming the intuition that controversial discussions exhibit a more complex structure. To provide additional insights on user interactions,
we consider also
the ratio of triangles in the reply graph \usersreplies
over the number of all possible triangles. 
 
\section{Experiments}
\label{sec:exp}

\subsection{Controversy Detection}

We evaluated different classifiers, including AdaBoost,
Logistic Regression, SVM and Random Forest,
and chose AdaBoost as it resulted in the best performance.
To show the relevance of detecting motifs to quantify controversy we compare the results with baseline graph-based features.
We analyzed the performance by the baseline graph-based features
and by using motif-based features (in addition and alone).

\begin{table}[b!]
\caption{Performance of the motif based classifier.}
\small
\label{tab:results} 
\begin{tabular}{lcccc}
\toprule
Filtering & {Accuracy} & {Precision} & {Recall} & {F-measure}\\ 
\midrule
\multicolumn{5}{c}{\scriptsize\textbf{\em Baseline}}\\
\midrule
$>$2 users 	& 0.76 & 0.79 & 0.81 & 0.80 \\
$>$3 users 	& 0.77 & 0.80 & 0.82 & 0.81 \\
$>$10 users 	& 0.78 & 0.81 & 0.83 & 0.82 \\
\midrule
\multicolumn{5}{c}{\scriptsize\textbf{\em Baseline $+$ dyadic motifs}}\\[-.2em]
\midrule
$>$2 users 	& 0.82 & 0.84 & 0.86 & 0.85\\
$>$3 users 	& 0.83 & 0.85 & 0.86 & 0.85\\
$>$10 users 	& \textbf{0.84} & \textbf{0.86} & {\bf 0.88} & {\bf 0.87}\\	
\midrule
\multicolumn{5}{c}{\scriptsize\textbf{\em Baseline $+$ dyadic and triadic motifs}}\\[-.2em]
\midrule
$>$2 users 	& 0.83 & 0.85 & 0.86 & 0.85\\
$>$3 users 	& 0.84 & 0.86 & 0.85 & 0.86\\
$>$10 users 	& {\bf 0.85} & {\bf 0.87} & {\bf 0.88} & {\bf 0.87}\\
\midrule
\multicolumn{5}{c}{\scriptsize\textbf{\em Dyadic motifs only}}\\[-.2em]
\midrule
$>$2 users 	& 0.75 & 0.77 & 0.82 & 0.80\\
$>$3 users 	& 0.75 & 0.77 & 0.82 & 0.80\\
$>$10 users 	& 0.77 & 0.79 & 0.84 & 0.82\\
\bottomrule
\end{tabular}
\end{table}

The baseline approach accuracy (with structural, propagation-based and temporal features) is above 75\%.
With the addition of dyadic motifs, all the performance figures are significantly improved.
The addition of triadic motifs leads to the best results,
but the improvement is only marginal because they are infrequent. 
The best results, highlighted in boldface, are statistically significant w.r.t.\ baseline features.
Using dyadic motifs alone, moreover, the accuracy of the model is comparable with the baseline.
We evaluated the importance of the features in the model: the first feature is the average inter-reply time:
when the discussion is polarized people tend to reply in a shorter time. 
The second most important feature is the
maximum relative degree.
The other features among the top-6 are dyadic motifs.
The most relevant being motif $A$: controversial threads create engagement among users
not being directly connected in the social network.
On the other hand, the fact that motif $C$ is not relevant
suggests that it is less likely to have controversial discussions among friends. 
Interestingly, dyadic patterns seem to be more relevant than propagation-based features.
We found also that it is not always appropriate to classify a reply tree
as controversial or not. This is because each reply may
generate unexpected reaction.
For instance, there may be sub-threads of controversy, within a non-controversial discussion.
To test this intuition, we analyzed the direct replies of the {\em origin} tweets 
that were classified as non-controversial.
This can be achieved easily as the proposed approach can be applied to any tweet
given its reply tree, or in this case, its reply sub-tree.
By applying the model discussed in the previous section,
we found that about 7\% of the direct-reply sub-trees of a non-controversial tweet are controversial.
Studying how the controversy related to a given hashtag evolves over time is an interesting task: for the sake of space we do not include further performed analyses on Twitter hashtags, but they confirm the efficacy of our approach in monitoring controversy over time. 

\section{Conclusion}
\label{sec:conclusion}

We proposed a novel language-independent approach based on local graph motifs
Such motifs correspond to different interaction patterns
among two users, which may be linked by a possibly reciprocal
reply action and by a possibly reciprocal friendship relationship. We proved on a benchmark Twitter dataset that such motifs are
more powerful in predicting controversy than other baseline frequently used
graph properties. 
We observed that in most cases controversy arise
when users participate to discussions beyond their social circles.
Finally, as the proposed motifs can be easily extracted from any reply tree
or sub-tree, we experimented with the use of such patterns
in monitoring the evolution of discussions and sub-discussions over time.
\section*{Acknowledgments}
\small{This work was partially supported by the EC H2020 Program INFRAIA-1-2014-2015 {\em SoBigData: Social Mining \& Big Data Ecosystem} (654024).}

\small
\bibliographystyle{aaai}
\bibliography{biblio}  \end{document}